\documentclass[useAMS,usenatbib]{mn2e} 
\usepackage{graphics} 
\usepackage{epsfig}
\usepackage{color}
\usepackage{ulem}
\usepackage[draft=false,colorlinks,dvipdfm=true,citecolor=blue,linkcolor=blue]{hyperref}
\usepackage{xspace}
\usepackage{txfonts}

\def\ni{\noindent}                                    

\newcommand{\Ha}{\ifmmode {\rm H}\alpha \else H$\alpha$\fi\xspace}
\newcommand{\Hb}{\ifmmode {\rm H}\beta \else H$\beta$\fi\xspace}
\newcommand{\Hba}{\ifmmode {\rm H}\beta^{\prime} \else H$\beta^{\prime}$\fi\xspace}
\newcommand{\Hg}{\ifmmode {\rm H}\gamma \else H$\gamma$\fi\xspace}
\newcommand{\Hd}{\ifmmode {\rm H}\delta \else H$\delta$\fi\xspace}
\newcommand{\hi}{H\,{\sc i}\xspace}

\newcommand{\Hii}{\ifmmode \rm{H}\,\textsc{ii} \else H\,{\sc ii}\fi}
\newcommand{\Nii}{[N\,{\sc ii}]$\lambda$6584}
\newcommand{\nii}{\ifmmode [\rm{N}\,\textsc{ii}] \else [N\,{\sc ii}]\fi\xspace}

\newcommand{\oi}{\ifmmode [\rm{O}\,\textsc{i}] \else [O\,{\sc i}]\fi\xspace}

\newcommand{\neiii}{\ifmmode [\rm{Ne}\,\textsc{iii}] \else [Ne\,{\sc iii}]\fi}

\newcommand{\hei}{\ifmmode \rm{He}\,\textsc{i} \else He\,{\sc i}\fi}
\newcommand{\oii}{\ifmmode [\rm{O}\,\textsc{ii}] \else [O\,{\sc ii}]\fi\xspace}
\newcommand{\Oiii}{[O\,{\sc iii}]$\lambda$5007}

\newcommand{\oiii}{\ifmmode [\rm{O}\,\textsc{iii}] \else [O\,{\sc iii}]\fi\xspace}

\newcommand{\sii}{\ifmmode [\rm{S}\,\textsc{ii}] \else [S\,{\sc ii}]\fi\xspace}
\newcommand{\siii}{\ifmmode [\rm{S}\,\textsc{iii}] \else [S\,{\sc iii}]\fi}

\newcommand\aj{{AJ}}
\newcommand\araa{{ARA\&A}}
\newcommand\apj{{ApJ}}
\newcommand\apjs{{ApJS}}
\newcommand\aap{{A\&A}}
\newcommand\aaps{{A\&AS}}
\newcommand\mnras{{MNRAS}}
\newcommand\pasp{{PASP}}

\defcitealias{CidFernandes.etal2010}{CF10}
\defcitealias{Stasinska.etal2006}{S06}
\defcitealias{Stasinska.etal2008}{S08}
\defcitealias{Bruzual.Charlot2003}{BC03}

\title[How to tell true from fake AGN?]{A comprehensive classification of galaxies in the SDSS: 
How to tell true from fake AGN? }

\author[Cid Fernandes et al]
         {R. Cid Fernandes$^{1}$,
	  G. Stasi\'nska $^{2}$,
 	  A. Mateus$^{1}$,
	  N. Vale Asari$^{1}$\\
	 $^{1}$Departamento de F\'{\i}sica - CFM - Universidade Federal de Santa Catarina,
	 Florian\'opolis, SC, Brazil\\
	 $^{2}$LUTH, Observatoire de Meudon, 92195 Meudon Cedex, France}

\begin{document}

\maketitle

\begin{abstract} 

  We use the $W_{\Ha}$ versus \nii/\Ha (WHAN) diagram introduced by us in previous work to provide a comprehensive emission-line classification of Sloan Digital Sky Survey galaxies. This classification is able to cope with the large population of weak line galaxies that do not appear in traditional diagrams due to a lack of some of the diagnostic lines. A further advantage of the WHAN diagram is to allow the differentiation between two very distinct classes that overlap in the LINER region of traditional diagnostic diagrams. These are galaxies hosting a weakly active galactic nucleus (wAGN) and ``retired galaxies'' (RGs), i.e. galaxies that have stopped forming stars and are ionized by their \emph{hot evolved low-mass stars} (HOLMES).

  A useful criterion to distinguish true from fake AGN (i.e. the RGs) is the value of $\xi$, which measures the ratio of the extinction-corrected \Ha\ luminosity with respect to the \Ha\ luminosity expected from photoionization by stellar populations older than $10^{8}$\,yr. We find that $\xi$ follows a markedly bimodal distribution, with a $\xi \gg 1$ population composed by systems undergoing star-formation and/or nuclear activity, and a peak at $\xi \sim 1$ corresponding to the prediction of the RG model.  We base our classification scheme not on $\xi$ but on a more readily available and model-independent quantity which provides an excellent observational proxy for $\xi$: the equivalent width of \Ha.  Based on the bimodal distribution of $W_{\Ha}$, we set the practical division between wAGN and RGs at $W_{\Ha}=3$ \AA.

Five classes of galaxies are identified within the WHAN diagram: 

\begin{itemize}
  \item Pure star forming galaxies: $\log \nii/\Ha\ < -0.4$ and  $W_{\Ha} > 3$ \AA
  \item Strong AGN (i.e., Seyferts): $\log \nii/\Ha\ > -0.4$ and  $W_{\Ha} > 6$ \AA
  \item Weak AGN: $\log \nii/\Ha\ > -0.4$ and  $W_{\Ha}$ between 3 and 6 \AA
  \item Retired galaxies (i.e., fake AGN): $W_{\Ha} < 3$ \AA
  \item Passive galaxies (actually, line-less galaxies): $W_{\Ha}$ and $W_{\nii} < 0.5$ \AA
\end{itemize}

A comparative analysis of star formation histories and of other physical and observational properties in these different classes of galaxies corroborates our proposed differentiation between RGs and weak AGN in the LINER-like family.  This analysis also shows similarities between strong and weak AGN on the one hand, and retired and passive galaxies on the other.

\end{abstract}


\section{Introduction}

 The goal of a classification scheme is to rationally organize a large collection of objects (animals, plants, books, galaxies) into a fixed number of smaller classes \emph{without leaving any object aside}, in order to help dealing with this large collection.  There are obviously many ways to classify objects, depending on what properties one is mostly interested in. For galaxies, one could, for example, be mainly interested in the morphology \citep{Hubble1936}, colours \citep{deVaucouleurs1960}, spectral features \citep{Morgan.Mayall1957} etc.  It is not only the considered object properties that change from one classification system to another, but also the philosophies underlying the classification process (hierarchical or non hierarchical, supervised or not supervised, \ldots) all of which have their advantages and drawbacks. Ideally, a classification scheme should be objective and allow easy incorporation of new objects into predefined categories. The last criterion, for instance, is not fulfilled by the classifications using principal component analysis \citep{Sodre.Cuevas1994, Connolly.etal1995}. Most importantly, the classification scheme must be useful, Hubble's famous tuning fork being a classical  example. Another one is  the emission-line classification scheme pioneered by \citet*{Baldwin.etal1981} (see also \citealt{Pastoriza1968}), which has been widely used over the last three decades.

 The advent of large surveys of galaxies providing extensive collections of homogeneous data sets not easily tractable with conventional methods fostered new classification schemes based on automatic procedures such as neural networks \citep*{Folkes.etal1996}, k-means cluster analysis of entire galaxy spectra \citep{SanchezAlmeida.etal2010}, or on human-eye analysis of galaxy morphology by hundreds of thousands of volunteers, as in the beautiful Galaxy Zoo project \citep{Lintott.etal2010}.

 Emission-line classifications of galaxies are among the easiest to carry out (once the emission line intensities are measured) and allow one to deal with such issues as star formation, chemical composition or nuclear activity.  They have, for example, helped to gain insight into the nature of warm infrared galaxies \citep{deGrijp.etal1992}, of luminous infrared galaxies \citep*{Smith.etal1998}, of galaxies in compact groups \citep{Coziol.etal1998} or into the connection between active galactic nuclei (AGN) and their host galaxies \citep{Kauffmann.etal2003c}. The most famous emission-line diagnostic diagram is that based on the \Oiii/\Hb\ vs \Nii/\Ha\ line ratios\footnote{In the remaining of this paper, \Oiii\ and \Nii\ will be denoted \oiii\ and \nii, respectively.} (dubbed the BPT diagram after \citealt{Baldwin.etal1981}). This diagram was built to pin down the main source of ionization in the spectra of extragalactic objects, and became one of the major tools for the classification and analysis of emission line galaxies in the Sloan Digital Sky Survey (SDSS, \citealt{York.etal2000}).
 
 As emphasized by \citet[CF10]{CidFernandes.etal2010}, the BPT diagram leaves unclassified a large proportion of emission line galaxies in the SDSS, due to quality requirements on 4 emission lines. The \citet{Kewley.etal2006} classification is even more demanding, as it requires the intensities of as many as 7 lines. To remedy this situation, \citetalias{CidFernandes.etal2010} have proposed a much more economic diagram that allows one to attain the same objectives using only two lines, \Ha\ and \nii, which are generally the most prominent in the spectra of galaxies. This is a diagram plotting the \Ha\ equivalent width versus \nii/\Ha, which was dubbed the EWH$\alpha$n2 diagram in that paper and which we will from now on refer to as the WHAN diagram, for simplicity \footnote{Our WHAN diagram is, in a way, very similar to the $W_{\nii}$ vs \nii/\Ha\  diagram of \cite{Coziol.etal1998}. However, the WHAN diagram is easier to interpret, since it concerns two physically independent quantities, $W_{\Ha}$ measuring the amount of ionizing photons absorbed by the gas relative to the stellar mass, while \nii/\Ha is a function of the nitrogen abundance, of the ionization state and temperature of the gas.}. The border lines between different classes of galaxies in the WHAN diagram were defined by optimal transpositions of the \citet{Kauffmann.etal2003c}, \citet[S06]{Stasinska.etal2006} and \citet{Kewley.etal2006} border lines onto the $W_{\Ha}$ vs \nii/\Ha\ plane.

 The categories of emission line galaxies considered by \citet{Kauffmann.etal2003c} and \citet{Kewley.etal2006} comprised star forming galaxies (SF), Seyferts, and LINERs\footnote{In those papers, LINERs should have rather been called ``galaxies with LINER-like spectra'', given that LINER is the abbreviation of "low ionization {\em nuclear} emission regions" (as originally introduced by \citealt{Heckman1980}), and the SDSS spectra cover much more than the nuclear regions of galaxies.} This latter category is supposed to comprise objects hosting low-level nuclear activity. However, as demonstrated by \citet[S08]{Stasinska.etal2008}, the LINER region in the BPT diagram also contains galaxies that have stopped forming stars and are actually ionized by the hot evolved low-mass stars (HOLMES) contained in them. This idea is not new. It was already put forward to explain the emission line spectrum of elliptical galaxies \citep{Binette.etal1994, Macchetto.etal1996} and even earlier in different contexts \citep{Hills1972, Terzian1974, Lyon1975, Sokolowski.Bland-Hawthorn1991}. This idea is recently gaining support \citep{Schawinski.etal2010, Masters.etal2010, Kaviraj2010}, especially with the detailed study of nearby galaxies \citep*{Sarzi.etal2010, Annibali.etal2010, Eracleous.etal2010}. As shown by \citetalias{CidFernandes.etal2010}, when taking into account the ``forgotten population'' of weak line galaxies, the proportion of such ``retired galaxies'' (RGs) increases dramatically. It is therefore essential to find ways to distinguish these RGs from truly ``active'' galaxies, i.e. galaxies which contain a (weak) AGN. This is one of the goals of the present paper.

Our second goal is to link the universe of emission line galaxies (ELGs) with the universe of galaxies without emission lines (often dubbed ``Passive Galaxies'', PG). We will see that RGs and passive galaxies are in fact very similar objects. 

With our WHAN diagram, we will thus be able to provide a comprehensive classification of all SDSS galaxies, including PGs.  

The paper is structured as follows: Section \ref{sec:Data} presents the samples of galaxies used to work out our classification and our {\sc starlight} data base from which the properties of galaxies that we use are extracted. Section \ref{sec:preliminary} lists the initial definition of galaxy classes: SF, Seyferts, LINERs and PGs.  Section \ref{sec:RGs}, the core of this paper, sorts out RGs from AGN among LINER-like galaxies and proposes a practical criterion to identify them. Section \ref{sec:Revised_EWHan2} describes our comprehensive classification of galaxies based on the WHAN diagram and leading to five classes: SF, strong AGN, weak AGN, RG and PG.  Section \ref{sec:SFHs} examines the star formation histories of galaxies within our new classification scheme, placing RGs in the context of the galaxy population in general. Section \ref{sec: distributions} discusses the distributions of a series of physical and observed properties in the different galaxy classes, to test the pertinence of our new classification.  Finally, Section \ref{sec:summary} summarizes our main results.

\section{Samples  and data products}
\label{sec:Data}

We draw from the nearly 700 thousand galaxies from the Main Galaxy Sample \citep{Strauss.etal2002} in the 7$^{\rm th}$ data release of the SDSS \citep{Abazajian.etal2009}.  Besides the SDSS data, we make use of a suite of {\sc starlight}-derived products available at www.starlight.ufsc.br, like stellar masses, stellar extinctions and the full star formation histories \citep{CidFernandes.etal2005}.  Emission line properties are measured after subtraction of the underlying stellar component determined with the {\sc starlight} fit, as described in \cite{Mateus.etal2006}.

Two samples are used throughout this work. The ``full sample'' (hereafter ``sample F'') is selected by requiring {\it (i)} a signal-to-noise ratio $\ge 10$ in the continuum around 4750 \AA, {\it (ii)} no faulty pixels within $\pm 15$ \AA\ of \Ha and \nii, {\it (iii)} $z < 0.17$ to eliminate the few objects at higher redshift which survive the previous criterion, and {(iv)} $z > 0.04$ to minimize apertures effects. In total, sample F contains 379330 galaxies.

Sample V is built from a volume limited sample satisfying $0.04 < z < 0.095$ and absolute r-band magnitude $M_r < -20.43$, to which we further apply criteria {\it (i)} and {\it (ii)} above. These latter cuts imply only a 6 percent reduction, with little loss of completeness at the faint end, so this sample is effectively limited in volume.  Sample V comprises 152749 objects.  By choosing $z < 0.095$ we eliminate cases where \Ha and \nii fall in a region of strong sky lines, so that criterion {\it (ii)} is fulfilled.

These samples fulfill the basic requirements for this work. Condition {\it (i)} ensures meaningful spectral fits, necessary for stellar population analysis and the production of templates to aid emission line measurements, while {\it (ii)} ensures that the detection of \Ha and \nii (the most important lines in this study) is not hampered by artifacts like bad pixels or sky residuals. At the same time, neither of these samples is biased for or against the presence of emission lines, which allows us to treat systems with strong, weak or no emission lines on an equal footing.

Sample F has the advantage of being more numerous and of extending to lower luminosities, while sample V is more adequate for demographic studies and unbiased comparisons among sub-samples.  The novelties in this work, namely the criterion to separate true from fake AGN and the resulting revision of conventional spectral taxonomy, are mainly related to the most luminous galaxies, so that, in practise, we could use either sample V or F. In what follows, unless stated otherwise, we use sample V.

\section{Preliminary galaxy classes}
\label{sec:preliminary}

We sub-divide galaxies into different classes according to their emission line properties. The first division is among galaxies with and ``without'' emission lines, denoted ELGs and PGs, respectively.

\subsection{Passive galaxies} 

PGs are defined as those with very weak or undetected emission lines.  Galaxies are deemed passive if the equivalent widths ($W_\lambda$) of {\em both} \Ha and \nii fall below 0.5 \AA, a criterion which applies to 19 (20)\% of sample V (F). Since these are the two strongest optical lines, this criterion automatically implies that other lines will be even weaker in general (e.g. \citealt{Brinchmann.etal2004}; \citetalias{CidFernandes.etal2010}).

In the literature one often finds PGs and ELGs defined in terms of limits on the signal to noise ratio ($SN_\lambda$) of emission lines \citep[e.g.][]{Miller.etal2003, Brinchmann.etal2004, Mateus.etal2006}.  We consider a $W_\lambda$-based criterion more appropriate, as it is based on a more direct measurement of the line strength and less explicitly dependent on the quality of the data.  In addition $W_\lambda$'s have an astrophysical meaning. We emphasize that the adopted limiting value of 0.5 \AA\ is not really meaningful. As will be seen later, none of the discussion presented in this paper would have changed if had we adopted a slightly different limit.  We also note in passing that the denomination ``Passive'' for ``line-less'' systems is adopted just for compatibility with current nomenclature, with no evolutionary connotation as the word would suggest.

\subsection{Emission Line Galaxies: SF, Seyferts and LINERs}

ELGs are defined as everything else, i.e., any source where either $W_{\Ha}$ or $W_{\nii}$ is $\ge 0.5$ \AA.  Notice that this stretches the definition of an ELG to its limit, since, in extreme cases, only one of \Ha or \nii is detected!  Sample V (F) has 124410 (303194) ELGs, 94 (92)\% of which have both lines stronger than 0.5 \AA.

\begin{figure}
\includegraphics[bb= 50 165 580 690,width=0.5\textwidth]{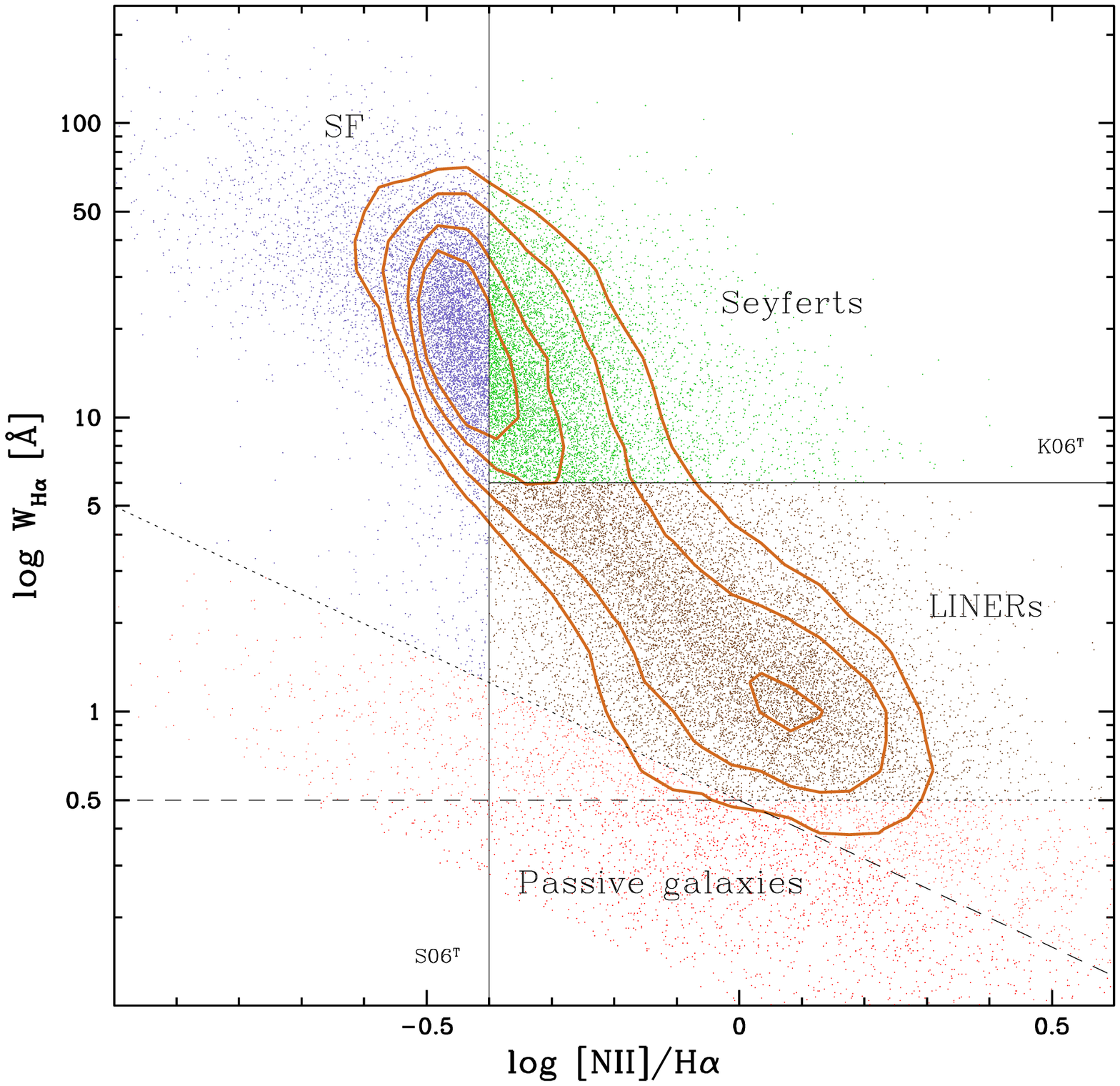}
\caption{The WHAN diagram, used to establish emission line classes.
  The line labelled S06$^T$ represents the optimal transposition of the
 \citetalias{Stasinska.etal2006} SF/AGN spectral classification scheme, while the line at
  $W_{\Ha} = 6$ \AA\ represents the transposed version of the \citet{Kewley.etal2006}
  Seyfert/LINER class division.  Dotted lines mark $W_{\Ha} = 0.5$ and
  $W_{\nii} = 0.5$ \AA\ limits, below which line measurements are
  uncertain.  Passive galaxies ($W_{\Ha}$ and $W_{\nii} < 0.5$ \AA,
  red dots) lie below the dashed line.  Dots correspond to sample F
  and contours to sample V.  Contours are drawn at number densities $=
  0.1$, 0.2, 0.4 and 0.6 of the peak value. As in other figures in
  this paper, {\em only 1 tenth of the  objects is plotted to avoid
    overcrowding. } }
\label{fig:EWHan2}
\end{figure}

ELGs are normally subdivided into SF and AGN-like classes on the basis of diagnostic diagrams involving at least 4 lines, like the BPT diagram (\oiii/\Hb versus \nii/\Ha). As shown in \citetalias{CidFernandes.etal2010}, the weakness of either or both of \Hb and \oiii implies that such traditional classification schemes leave large numbers of ELGs unclassified. This problem is particularly severe in the right wing of the BPT diagram, where AGN-like systems reside, but well over half of the galaxies have unusable \Hb and/or \oiii fluxes.  In that same paper we introduced alternative diagnostic diagrams able to cope with galaxies with weak or undetected \Hb and/or \oiii, the most economic of which is $W_{\Ha}$ versus \nii/\Ha, hereafter the WHAN diagram.  We base all our classifications on this diagram, which, unlike any other, allows us to classify {\em all} galaxies.

The WHAN diagram is shown in Fig.\ \ref{fig:EWHan2}.  The vertical line at $\log \nii/\Ha = -0.40$ corresponds to the optimal transposition of the \citetalias{Stasinska.etal2006} BPT-based SF/AGN division, designed to differentiate sources where star formation provides all ionizing photons from those where a harder ionizing spectrum is required. Similarly, the division at $W_{\Ha} = 6$ \AA\ represents an optimal transposition of the \citet{Kewley.etal2006} division between Seyferts and LINERs on the basis of 7 emission lines. Unlike \citet{Kewley.etal2006} and other studies (e.g., \citealt{Kauffmann.etal2003c}; \citealt{Mateus.etal2006}), we do not define a ``SF+AGN composite'' category, and so our Seyfert/LINER border line goes all the way to the SF/AGN frontier.  (The reader is referred to \citetalias{CidFernandes.etal2010} for a discussion of the pros and cons of this diagram, as well as for a discussion of why we prefer not to define a composite category.)  Sample V splits into percentage fractions of 23, 21 and 37 SF, Seyferts and LINERs, respectively. PGs account for the remaining 19\%. The corresponding fractions for sample F are very similar: 25, 19, 36 and 20\%.

The dotted lines in Fig.\ \ref{fig:EWHan2} correspond to $\min(W_{\Ha},W_{\nii}) = 0.5$ \AA. Points below these lines (painted in orange or red) are thus very uncertain\footnote{ Our line fitting code, described in \citet{Mateus.etal2006}, only accepts fits with $W_\lambda \ge 0.1$ \AA, which produces the empty triangle in the bottom left part of the plot.}, but are nevertheless plotted to remind the reader of the existence of such objects. There is indeed some advantage in including line-less galaxies in an emission line classification scheme since the absence of emission lines is also an information!  The region below the dotted lines includes both ELGs with only one line $\ge 0.5$ \AA, as well as PGs (below the dashed lines, in red) for which we have \Ha and \nii fluxes in our database.

\section{Identifying retired galaxies}
\label{sec:RGs}

Ideally, splitting ELGs into SF and AGN would separate systems whose emission lines are powered exclusively by young stars from those where accretion onto a central super-massive black hole dominates (or at least contributes to) the ionizing radiation field.  However, as reviewed in the introduction, this binary scheme does not account for a third phenomenon: photoionization by HOLMES.  These Retired Galaxies (RGs) are numerous in large-scale surveys \citepalias{Stasinska.etal2008}, where they occupy LINER-like locations in the commonly used diagnostic diagrams, and are thus erroneously associated to non-stellar activity.

This section investigates ways of identifying RGs.  We start with a theoretically motivated definition, but end up with a very simple empirical criterion to diagnose RGs among ELGs.

\subsection{Ionizing photons budget: The $\xi$ ratio}
\label{sec:XI_model}

By definition, RGs are systems whose old stellar populations suffice to explain the observed emission line properties.  Since we already know that nebulae photoionized by HOLMES cover most regions in intensity ratios diagrams \citepalias{Stasinska.etal2008}, we shall adopt an operationally simple energy-budget criterion and define RGs as {\em galaxies whose old stellar populations are capable of accounting for all of the observed \Ha emission}.

Let $L^{\rm int}_{\Ha}$ be the intrinsic (extinction corrected) \Ha luminosity of a galaxy, and $L^{\rm exp}_{\Ha}(t>10^8 {\rm yr})$ denote the luminosity expected from photoionization by populations older than $10^8$ yr.  The ratio

\begin{equation}
\label{eq:xi}
\xi = \frac{L^{\rm int}_{\Ha}} {L^{\rm exp}_{\Ha}(t>10^8 {\rm yr})}
\end{equation}

\ni should thus be able to tell if a galaxy is retired ($\xi \le 1$) or whether some extra source of ionizing photons is present.  Given the orders of magnitude difference between the ionizing photon rates of young and old populations, even tiny amounts of ongoing star formation suffice to swamp the ionizing field produced by HOLMES, leading to $\xi \gg 1$ and to a SF-like emission-line pattern due to the softer ionizing spectrum. Similarly, $\xi$ will also grow above 1 in the presence of an AGN.  In fact, as long as a galaxy has enough gas to absorb all $h\nu > 13.6$ eV photons produced within it, $\xi$ should never drop below 1.

This way of identifying RGs was previously employed by \citetalias{Stasinska.etal2008}. In what follows we open up the details and discuss the caveats related to the estimation of $\xi$.

\subsection{Computation of the expected \Ha luminosity}
\label{sec:LHa_expected}

Neglecting escape and extinction of Lyman continuum radiation, the denominator in eq.\ \ref{eq:xi} becomes

\begin{equation}
L^{\rm exp}_{\Ha}(t>10^8 {\rm yr}) =  \frac{h\nu_{\Ha}}{f_{\Ha}}
\times Q_H^{\rm exp}(t>10^8 {\rm yr})
\end{equation}

\ni where $f_{\Ha} = 2.226$ for case B hydrogen recombination, and $Q_H^{\rm exp}$ is the expected ionizing photon rate. The {\sc starlight} spectral decomposition in terms of a base of simple stellar populations (SSPs) allows us to estimate $Q_H^{\rm exp}$ for each of our galaxies.
The code returns the mass associated to each of $N_\star$ SSPs of different ages ($t$) and metallicities ($Z$), so that the expected flux of ionizing photons from the populations of interest here is simply

\begin{equation}
\label{eq:Q_exp}
Q_H^{\rm exp}(t>10^8 {\rm yr}) = M_\star \sum_{j ; t_j > 10^8 yr} \mu_j q_{H,j}
\end{equation}

\ni where $M_\star$ is the total stellar mass formed, $\vec{\mu}$ is the mass-fraction population vector, and $q_{{\rm H},j} = q_{\rm H}(t_j,Z_j)$ is the number of H-ionizing photons emitted per unit time and initial mass for the $j^{\rm th}$ SSP.  Since the expected \Ha luminosity is to be compared to the one observed within the 3\arcsec aperture of the SDSS spectroscopic fiber, the stellar mass in this equation corresponds to that inside the fiber (i.e., we do not extrapolate to account for the whole galaxy).  In fact, all our analysis up to Section \ref{sec:SFHs} uses within-the-fiber data only.

The {\sc starlight} fits used in this work employ the same base of SSPs from \citet[hereafter BC03]{Bruzual.Charlot2003} described by \citet{Mateus.etal2006}. Naturally, the spectral fits are carried out in the optical range, but the \citetalias{Bruzual.Charlot2003} models also provide predictions for the ionizing energies.  This is a key element of our analysis, so it is worth to open a parenthesis to examine these predictions more closely.

\subsection{SSP models beyond 1 Rydberg}
\label{sec:SSPs}

\begin{figure*}
\includegraphics[bb= 35 160 575 540,width=\textwidth]{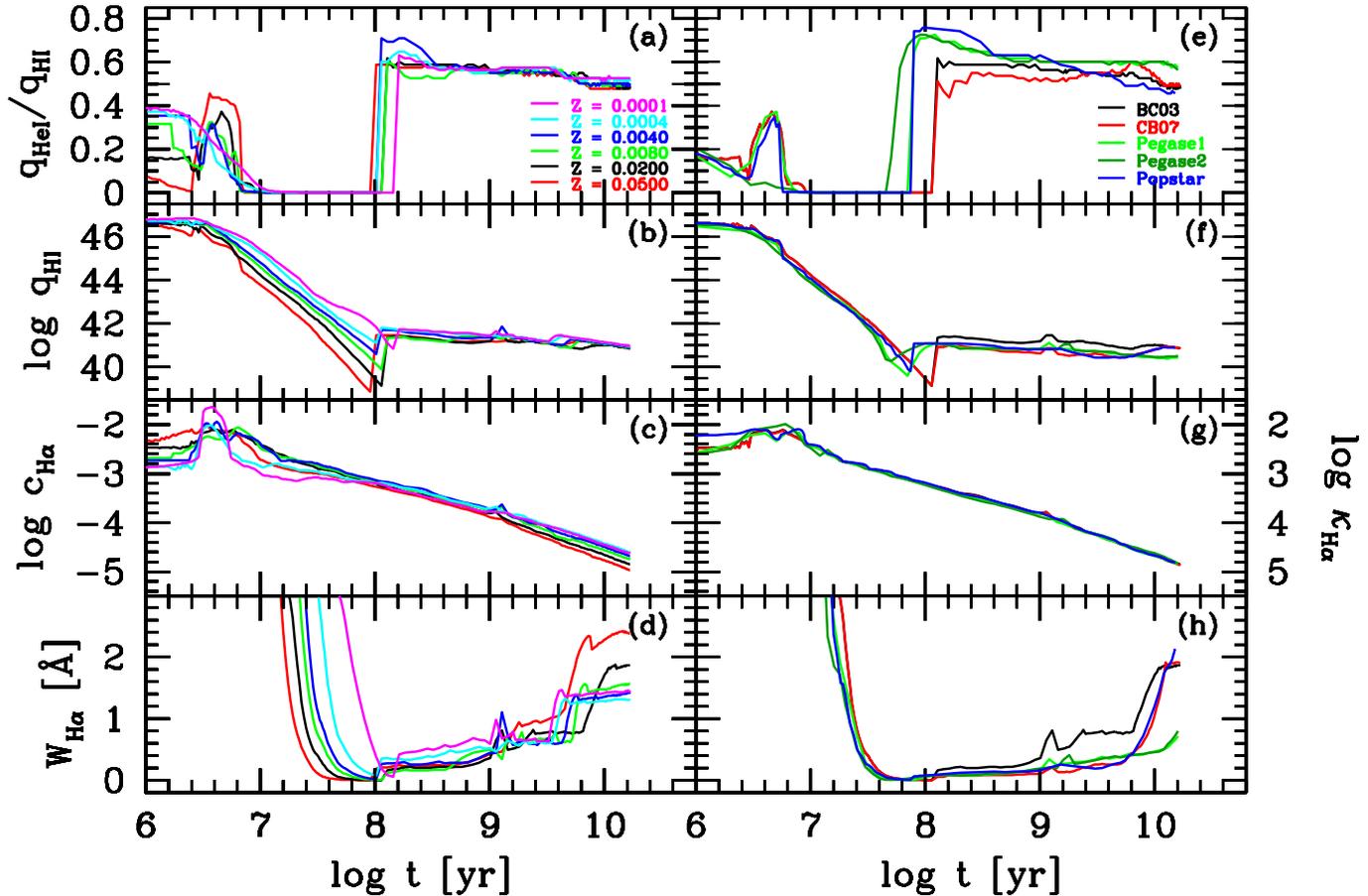}
\caption{Evolution of the ratio between the \hei\ and \hi ionizing
  photon rates (panel a), \hi-ionizing  photon rate (b, in units of $s^{-1} M_\odot^{-1}$), \Ha continuum (c, in
  units of $L_\odot  {\rm \AA}^{-1} M_\odot^{-1}$), and
  \Ha equivalent width (d) for simple stellar populations in the
  \citetalias{Bruzual.Charlot2003} models of metallicities between
  $Z_\star = 0.005$ and $2.5 Z_\odot$.  Panels on the right are
  $Z_\odot$ models from 5 different sources (including
  \citetalias{Bruzual.Charlot2003} for comparison). See text for
  details. The right axis scale in panels c and g gives the mass to
  light ratio in the \Ha continuum: $\kappa_{\Ha} = c_{\Ha}^{-1}$ (in units of 
$M_\odot L_\odot^{-1}   {\rm \AA}$).
}
\label{fig:SSPs}
\end{figure*}

Fig.\ \ref{fig:SSPs} shows the evolution of SSP properties of interest to this work.  Panels on the left (a--d) correspond to \citetalias{Bruzual.Charlot2003} models of 6 different metallicities used in our analysis. These are discussed first.

Fig.\ \ref{fig:SSPs}a shows the evolution of the hardness of the ionizing spectrum, as measured by the ratio of HeI to HI ionizing photons.  The difference between the soft ionization field produced by OB stars ($t < 10^7$ yr) and that produced by HOLMES is clearly visible.  Panel b shows the specific ionizing photon rate, $q_{\rm H}(t,Z)$, the most relevant quantity for our analysis. One sees the well known drop by $\sim 5$ orders of magnitude in $q_{\rm H}$ from the HII region phase to the time when the HOLMES regime sets in at $10^8$ yr and $q_{\rm H}$ approximately stabilizes at a level of the order of $10^{41}\,$s$^{-1}\,$M$_\odot^{-1}$. The specific continuum around \Ha, $c_{\Ha}$ (equivalent to the inverse of the mass to light ratio, $\kappa_{\Ha}$), however, keeps decreasing with time (panel c) and therefore the predicted $W_{\Ha}$ (bottom panel) goes through a minimum at $10^8$ yr and grows to 1.5--2.5 \AA\ at $t \ga 10^{10}$ yr. This is enough to explain a significant proportion of the ELGs in the LINER zone of Fig.\ \ref{fig:EWHan2}.

The right column of panels Fig.\ \ref{fig:SSPs} shows the predictions obtained for other evolutionary synthesis models at solar metallicity and \citet{Chabrier2003} initial mass function (IMF, the same one used for the BC03 models on the left panels).  Red lines show results for a preliminary version of Charlot \& Bruzual models. Blue lines are used for Popstar models \citep*{Molla.etal2009} with \citet*{Lejeune.etal1997} library and NLTE models by \citet{Rauch2003} and ``Padova 1994'' tracks (version updated by \citealt{Molla.GarciaVargas2000}). Green lines represent Pegase models \citep{Fioc.Rocca-Volmerange1997} with their 200 \AA--10 $\mu$m stellar library (see their section 2.2.1 for more details). Two sets of evolutionary tracks are shown for Pegase: Light green is used for \citet{Bressan.etal1993} tracks up to TP-AGB, dark green shows the set for \citet{Schaller.etal1992} and \citet{Charbonnel.etal1996} tracks. Both sets use \citet{Schoenberner1983} and \citet{Bloecker1995} for the PAGB.  For reference, the \citetalias{Bruzual.Charlot2003} predictions for $Z_{\odot}$ are repeated (black lines).

Despite a general qualitative agreement, there are significant quantitative differences. For ages above $10^{8}$\,yr, the hardness ratio varies from 10 to 20\%, while $q_{\rm H}$ differs by up to 1 dex from one model to another. Those differences could be due to differences in stellar tracks, model atmospheres and even interpolation techniques. For $W_{\Ha}$, the greatest difference is above $10^{9}$\,yr for the \citetalias{Bruzual.Charlot2003} models, which are $\sim 0.5$ \AA\ larger than the others.  We have also compared different IMFs (e.g., \citealt{Salpeter1955} and \citealt{Kroupa2001}). For a given metallicity in a model, different IMFs result in variations from 0.2 up to 0.5 dex in $q_{\rm H}$ and $c_{\Ha}$, which is typically of the same order of variations for different models with the same IMF.

For the sake of internal consistency with our {\sc starlight} analysis of SDSS galaxies, in what follows we adopt the predictions from \citetalias{Bruzual.Charlot2003}, but this quick compilation illustrates that there are still plenty of uncertainties in model predictions for the Lyman continuum of old stellar populations. Naturally, these uncertainties propagate to our estimate of $\xi$.  Fortunately, there is an empirical alternative to circumvent the uncertainties related to this choice (Section \ref{sec:WHaDefinitionOfRGs}).

\subsection{Extinction correction}
\label{sec:ExtinctionCorrection}

Let us now turn to the numerator of eq.\ (\ref{eq:xi}). The caveat in this case is that a nebular extinction is needed to correct $L_{\Ha}^{\rm obs}$ to $L_{\Ha}^{\rm int}$, which in turn requires good measurements of both \Ha and \Hb. While this is not a problem for ELGs in the SF and Seyfert categories, most of the LINER-like systems do not comply with this requirement because of their feeble \Hb.

One way to circumvent this difficulty is to estimate \Ha/\Hb in terms of something else. To do this, we first select LINERs with reliable \Ha/\Hb, and then calibrate the relation between nebular ($A_V^{\rm neb}$) and stellar ($A_V^\star$) extinctions (as done by \citealt{Asari.etal2007} for SF galaxies), such that $A_V^\star$ can be used to estimate $A_V^{\rm neb}$. In practice, we impose $W_{\Ha}$ and $W_{\Hb} > 0.5$ \AA, and $\Ha/\Hb > 3$ (such that $A_V^{\rm neb} > 0$)\footnote{For the \citet{Cardelli.etal1989} law used in our analysis, $A_V^{\rm neb} = 7.215 \log \frac{ (\Ha/\Hb)_{\rm obs} }{ (\Ha/\Hb)_{\rm int } }$. We adopt an intrinsic Balmer decrement of 3, suitable for LINERs \citepalias{Stasinska.etal2008}.}, which leaves us with only $\sim 30\%$ of the full list of LINERs.  Nebular and stellar extinctions are indeed correlated for this sub-sample (Spearman rank correlation coefficient $R_S = 0.38$), albeit with substantial scatter.  A fit through the median relation yields $A_V^{\rm neb} = 1.56 A_V^\star + 0.53$, with typical residuals of $\sim 0.5$ mag.  Objects used in this calibration have $A_V^{\rm neb} = 1.0 \pm 0.5$ (median $\pm$ semi inter-quartile range), while for LINERs as a whole the above relation leads to $A_V^{\rm neb} = 0.8 \pm 0.2$.  This method most likely {\em overestimates} $A_V^{\rm neb}$, as there is a trend of increasing $A_V^{\rm neb} $ with increasing $W_{\Ha}$, and \Ha is much weaker in the full sample than in the sub-set used to calibrate the $A_V^{\rm neb}(A_V^\star)$ relation ($W_{\Ha} = 1.5$ versus 3.7 \AA\ in the median, respectively).

Another alternative is to apply the extinction correction when the data allow ($W_{\Ha}$ and $W_{\Hb} > 0.5$ \AA, and $\Ha/\Hb > 3$), and assume $A_V^{\rm neb} = 0$ otherwise.

These methods lead to estimates of $\xi$ which should bracket the correct solution.  Since $A_V^{\rm neb} = 1$ implies a 0.32 dex increase in $L_{\Ha}^{\rm int}$, for the typical inferred $A_V^{\rm neb}$ values these two methods should differ by $\sim$ 0.2--0.3 dex in $\xi$.

\subsection{Results: The bimodal distribution of $\xi$}
\label{sec:xi_Hists}

\begin{figure}
\includegraphics[bb= 50 170 430 690,width=0.5\textwidth]{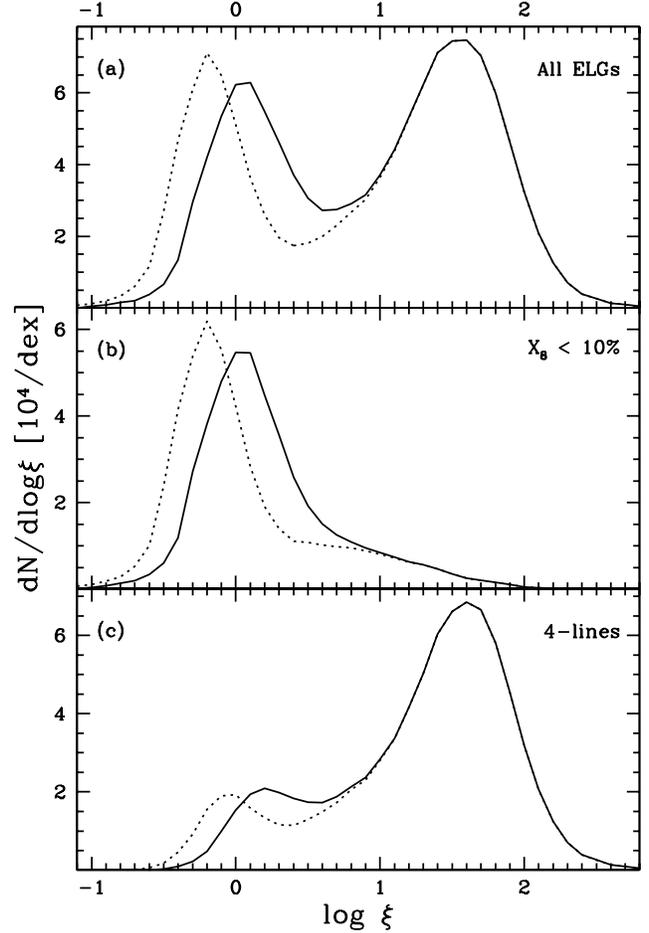}
\caption{ Histograms of the ratio between the intrinsic \Ha luminosity
  $L_{\Ha}$ to that predicted to originate from photoionization by HOLMES
  (equation \ref{eq:xi}) for ELGs in our volume limited sample,
  and for two choices of extinction correction for objects with
  unreliable \Ha/\Hb: estimate $A_V^{\rm neb}$ from $A_V^\star$ (solid
  lines), and no-correction (dotted).  Panel (a) includes all sources,
  while in (b) we select galaxies whose populations younger than
  $10^8$ yr contribute less than $X_8 = 10\%$ to the continuum at $\lambda = 4020$
  \AA. Panel (c) shows the $\xi$ distribution which results from
  requiring convincing detections of \Hb and \oiii, as well as \Ha and
  \nii.}
\label{fig:xi_Hists}
\end{figure}

The $\xi$ values derived range from $\sim$ 0.1 to 1000. This is illustrated in Fig.\ \ref{fig:xi_Hists}a, which shows the distribution of $\xi$ values for all ELGs in sample V. The distribution is strongly {\em bimodal}, with a high $\xi$ peak at $\sim 30$ and a lower one centred close to $\xi = 1$.

The two histograms in Fig.\ \ref{fig:xi_Hists}a differ in the recipe to correct $L_{\Ha}^{\rm obs}$ for extinction when \Ha/\Hb is unreliable. The dotted curve is for no correction, whereas the solid one is for the $A_V^{\rm neb} (A_V^\star)$ method discussed above. The curves are identical for $\xi > 10$ because these are Seyfert and SF galaxies, with \Ha and \Hb fluxes good enough to apply the standard correction.  In the low $\xi$ peak the two methods differ by just 0.2 dex on the median, as anticipated. Considering the smallness of this difference and that these recipes are likely to bracket the range of possibilities, we henceforth use the average $\xi$, whose distribution peaks at $\xi \sim 1$.

Overall, the low $\xi$ population is unbelievably close to the exact prediction for RGs: $\xi = 1$! But are these galaxies really retired in the sense of not having formed stars for a long while?  To answer this question, Fig.\ \ref{fig:xi_Hists}b shows the histograms obtained selecting only galaxies whose populations younger than $10^8$ yr contributes no more than $X_8 = 10$\% of the light at 4020 \AA.\footnote{To have an idea of what this means, $X_8 < 10$\% corresponds to specific star formation rates $\la 2.5 \times 10^{-12}$ yr$^{-1}$, or, equivalently, to less than 0.025 percent of the stellar mass having formed in the past $10^8$ yr.  In fact, for light levels as low as $X_8 = 10\%$ the spectral synthesis analysis is affected both by its own limitations and by side-effects of incompleteness in evolutionary synthesis models, such that the true contribution from young stars cannot be reliably estimated (see Section \ref{sec:SFHs}).}  Clearly, the low $\xi$ peak is made up of galaxies which have formed little or no stars in the past $10^8$ yr, in full agreement with the RG scenario of \citetalias{Stasinska.etal2008}.

The width of the RG peak in the $\xi$ distribution (as measured from the one-sided dispersion of its lower half, and assuming that the distribution is symmetric) is $\sigma_{\log\xi} \sim 0.24$ dex for the full ELG sample or 0.13 dex eliminating those with $W_{\Ha} < 0.5$ \AA.  From the considerations in the above sections, one would hardly expect $\xi$ to be more precise than a factor of 2, so that this observed dispersion can be accounted for by uncertainties in the data and the analysis.

While theory says that RGs should have $\xi \le 1$, the {\em observed} bimodality in the $\xi$ distribution suggests that $\xi < 2$ or 3 is a more effective way of separating RGs from the rest of the ELG population (Fig.\ \ref{fig:xi_Hists}a).  About 1/3 of the full ELG sample has $\xi < 2$.  The reason why this huge population of RGs has received little attention to date is because standard selection criteria tend to exclude them from emission line studies. Fig.\ \ref{fig:xi_Hists}c illustrates how the distribution of $\xi$ changes by imposing a $W_\lambda > 0.5$ \AA\ quality control for the 4 lines in the BPT diagram: \Hb, \oiii, \Ha and \nii. This reduces the ELG sample as a whole by $\sim 1/3$, but the fraction of $\xi < 2$ sources decreases by a factor of 3.  The bimodality is still present, but the low $\xi$ peak is heavily suppressed, so that RGs are the main victims of this cut.  With the WHAN diagram we are able to rescue RGs and weak line galaxies in general, eliminating this bias and thus drawing a much more complete view of ELGs in the local Universe.

\subsection{Results: $\xi$ versus \nii/\Ha}

\begin{figure}
\includegraphics[bb= 50 165 580 690,width=0.5\textwidth]{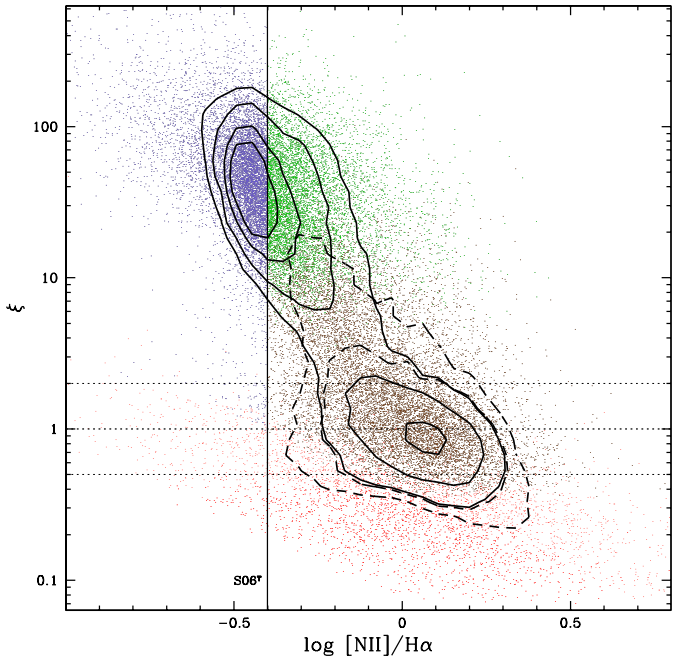}
\caption{ELGs in the $\xi$ versus \nii/\Ha diagram, color coding PGs,
  LINERS, Seyferts and SF, as in Fig.~\ref{fig:EWHan2}.  Also as in
  Fig.~\ref{fig:EWHan2}, gray points are sources with either or both
  of \Ha and \nii weaker than 0.5 \AA. Solid contours are for the full
  sample V, while the dashed ones are for galaxies with less than $X_8
  = 10\%$ of their flux at $\lambda = 4020$ \AA\ due to populations
  younger than $10^8$ yr (for clarity, only the two outer contours are
  plotted in this case). Dotted horizontal lines mark a range of
  $\times 2$ around the prediction of the retired galaxy model ($\xi = 1$).}
\label{fig:xi_X_N2Ha}
\end{figure}

Fig.~\ref{fig:xi_X_N2Ha} shows the results of this analysis in a $\xi$ versus \nii/\Ha diagram, color coding SF, Seyferts and LINERs according to their location on the WHAN diagram (Fig.\ \ref{fig:EWHan2}).  The plot shows that SF, Seyferts and LINERs are well separated in terms of $\xi$ and \nii/\Ha, with SF and Seyferts being responsible for the $\xi \gg 1$ peak and LINERs for the $\xi \sim 1$ RG-like population.  The plot also illustrates how dangerous it is to define a galaxy as an AGN on the basis of \nii/\Ha alone.  \citet{Miller.etal2003}, for instance, define ``2-lines AGN'' as sources with $\log \nii/\Ha > -0.2$ (similar to the ``low S/N AGN'' in \citealt{Brinchmann.etal2004}), while our analysis suggests that most of this population does not require an energetically significant non-stellar source of ionizing photons.

The dashed contours in Fig.~\ref{fig:xi_X_N2Ha} show the location of ELGs with $X_8 < 10\%$. This clarifies that the bulk of the population of large $\xi$ sources among galaxies with a predominantly old stellar population (seen in the middle panel of Fig.\ \ref{fig:xi_Hists}) is composed of LINERs approaching the Seyferts zone. Further considering that the $X_8$ cut eliminates sources with significant on-going star formation, this strongly indicates that these sources have a non-stellar source powering most of its line emission (i.e., they are AGN in old hosts).

We note in passing a tiny population of galaxies with small $\xi$ (i.e.  consistent with our RG model) but located to the left of the \citetalias{Stasinska.etal2006} line in Fig.~\ref{fig:xi_X_N2Ha}. These could correspond low-mass, low-metallicity galaxies in a quiescent phase between bursts of star formation, such as the galaxies discussed by \citet{SanchezAlmeida.etal2008}.

\subsection{$\xi$ versus $W_{\Ha}$}
\label{sec:xi_X_WHa}

\begin{figure}
\includegraphics[bb= 30 165 300 630,width=0.5\textwidth]{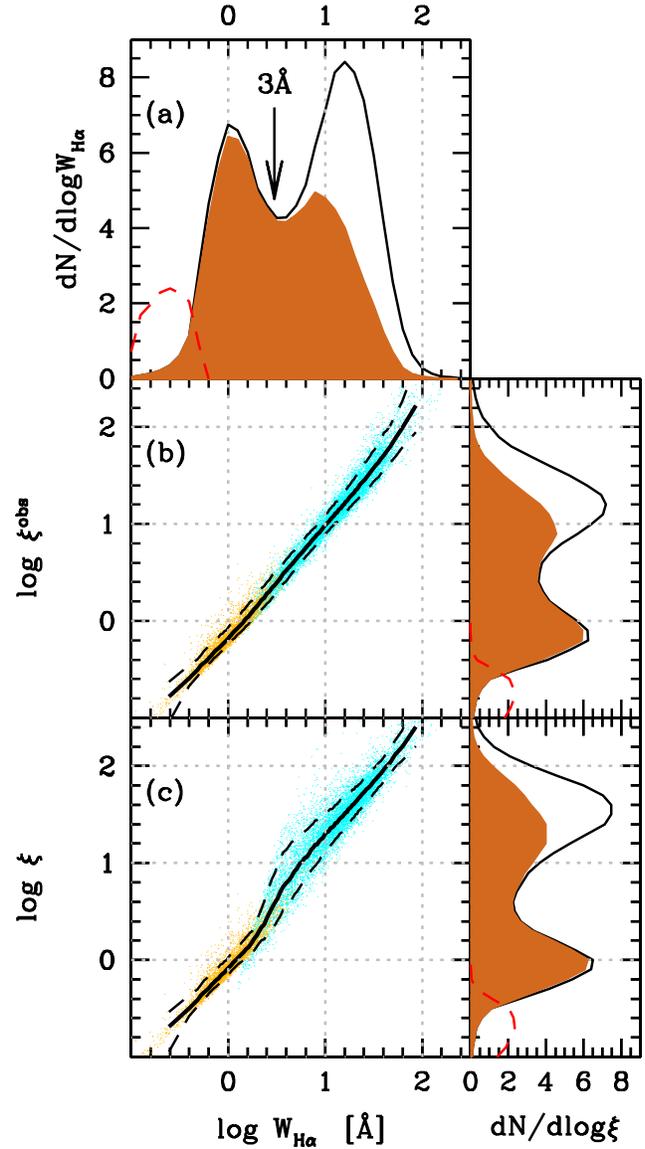}
\caption{Relation between $\xi$ and $W_{\Ha}$.  The bottom panel (c)
  shows the average $\xi$ value discussed in Section
  \ref{sec:ExtinctionCorrection}, while in the middle one (b) no
  extinction correction correction was applied to $L_{\Ha}$.  Points
  in cyan (orange) correspond to galaxies where \Ha/\Hb is  (is not)  reliable.
  The black solid line panels b and c shows the median relation, while
  the dashed lines represent the 10 and 90\% percentiles . In all
  projected histograms the black line represents all ELGs, while the
  filled region represents those with $\log \nii/\Ha > -0.4$ (i.e., to
  the right of the S06$^T$ line in the WHAN diagram), and the
  dashed red line indicates PGs for which a (necessarily small and
  uncertain) \Ha flux is available. The arrow in the $W_{\Ha}$
  distribution marks the proposed limit to separate RGs from other
  ELGs.}
\label{fig:xi_X_WHa}
\end{figure}

An important aspect of Fig.~\ref{fig:xi_X_N2Ha} is its striking similarity to the WHAN diagram of Fig.\ \ref{fig:EWHan2}. This happens because $\xi$ and $W_{\Ha}$ are very closely related.  The equivalent width of \Ha is defined as $W_{\Ha} = L_{\Ha}^{\rm obs} / C_{\Ha}^{\rm obs}$, where the denominator is the observed continuum luminosity around \Ha\footnote{$C_{\Ha}$ is not to be confused with the continuum luminosity per unit mass, $c_{\Ha}$, introduced in Section \ref{sec:SSPs}.  We use capital letters for extensive quantities (like $C_{\Ha}$ and $Q_H$), and lower case for specific ones ($c_{\Ha}$ and $q_H$). }. Hence, extinction corrections aside, $W_{\Ha}$ and $\xi$ (equation \ref{eq:xi}) have the same numerator.

Because $q_{H}(t,Z)$ varies little with $t$ or $Z$ after $10^8$ yr (Fig.\ \ref{fig:SSPs}b and f), and since most of the mass is always in the old populations (over 99\% for our samples), $Q_H^{\rm exp}(t > 10^8 {\rm yr})$ in equation (\ref{eq:Q_exp}) becomes $\sim M_\star q_{\rm H}^{\rm old}$, where $q_{\rm H}^{\rm old}$, the specific ionizing flux due to HOLMES, is of the order of $10^{41}$ s$^{-1}$M$_\odot^{-1}$. $M_\star$ can be written in terms of the intrinsic continuum at \Ha and the galaxy's luminosity weighted mass-to-light ratio: $M_\star = C_{\Ha}^{\rm int} \langle \kappa_{\Ha} \rangle_L$, where $\kappa_{\Ha} = c_{\Ha}^{-1} $ was previously discussed in Section \ref{sec:SSPs}. While $\kappa_{\Ha}$ spans nearly 3 decades for SSPs between $10^6$ and $10^{10}$ yr (Fig.\ \ref{fig:SSPs}c), galaxies are not SSPs, and young (low $\kappa_{\Ha}$) populations rarely make a substantial contribution to the \Ha continuum, so that the luminosity weighted $\kappa_{\Ha}$ values are of the same order for most sources. For the V sample, for instance, $\log \langle \kappa_{\Ha} \rangle_L = 4.56 \pm 0.14$ (median $\pm$ semi-interquartile range), while sub-samples of SF, Seyferts, LINERs and PGs have $4.35 \pm 0.12$, $4.47 \pm 0.08$, $4.65 \pm 0.08$ and $4.70 \pm 0.06$, respectively (in units of M$_\odot\,$L$_\odot^{-1}\,$\AA). As a result, $L_{\Ha}^{\exp}$ is directly proportional to the \Ha continuum, with little scatter due to variations in the star formation histories, and thus the denominators of $\xi$ and $W_{\Ha}$ essentially measure the same thing.

With these approximations, one finds that

\begin{equation}
\label{eq:xi_X_WHa}
  \xi =  \frac{ W_{\Ha} }{ W_{\Ha}^{\rm old} }  10^{0.4 (A_{\Ha}^{\rm neb} -  A_{\Ha}^\star)} 
\end{equation}

\ni where the $A_{\Ha}^{\rm neb} - A_{\Ha}^\star$ term comes from the fact that $\xi$ accounts for possible differences in the nebular and stellar extinctions, whereas $W_{\Ha}$ does not, and

\begin{equation}
  W_{\Ha}^{\rm old} \equiv  \frac{ h\nu_{\Ha} }{ f_{\Ha} }
  \langle \kappa_{\Ha}  \rangle_L q_{\rm H}^{\rm old}
  = 0.355 {\rm \AA}
  \left[ \frac{ \langle \kappa_{\Ha} \rangle_L }{ 10^4 } \right]
  \left[ \frac{ q_{\rm H}^{\rm old} }{ 10^{41} } \right]
\end{equation}

\ni is the expected equivalent width for an old population (for $\langle \kappa_{\Ha} \rangle_L$ in units of M$_\odot\,$L$_\odot^{-1}\,$\AA, and $q_{\rm H}^{\rm old}$ in s$^{-1}\,$M$_\odot^{-1}$).

Fig.\ \ref{fig:xi_X_WHa} shows the observed relation between $\xi$ and $W_{\Ha}$.  The bottom panel (c) shows the dust corrected $\xi$ value.  Points in cyan and orange correspond to galaxies where \Ha/\Hb is or is not reliable, respectively. The correlation is very tight, as indicated by the 10 and 90\% percentile lines (in dashed).  However, the differential extinction correction to $\xi$ creates a hump starting around $W_{\Ha} \sim 2$ \AA, coinciding with the transition between \Ha/\Hb-based corrections and those based on other methods (see Section \ref{sec:ExtinctionCorrection}).

Fig.\ \ref{fig:xi_X_WHa}b shows the same relation assuming that \Ha and its continuum are equally affected by dust ($A_V^{\rm neb} = A_V^\star$), so that the last term in equation (\ref{eq:xi_X_WHa}) disappears, and so does the hump in Fig.\ \ref{fig:xi_X_WHa}c.  The relation is linear except at high $W_{\Ha}$, where it steepens because of the decrease in $\langle \kappa_{\Ha} \rangle_L$ in galaxies with lots of star-formation. A fit through the median curve in its low $W_{\Ha}$ end yields $\log \xi^{\rm obs} = 1.07 \log W_{\Ha} - 0.17$. Forcing a linear fit leads to $W_{\Ha}^{\rm old} = 0.67$ \AA.  For comparison, this value of $W_{\Ha}^{\rm old}$ is reached for ages of a few Gyr for the \citetalias{Bruzual.Charlot2003} SSPs, or closer to 10 Gyr for other models (bottom panels in Fig.\ \ref{fig:SSPs}).


\subsection{Retired Galaxies: An empirical diagnostic}
\label{sec:WHaDefinitionOfRGs}

This section started with the promise of finding a diagnostic to identify RGs. The $\xi$ ratio was introduced with this purpose, and we have seen that it is indeed capable of capturing in a single number the essence of RGs. However, there are caveats related to this index. Firstly, uncertainties related to the extinction correction affect $\xi$ at a level of $\pm 0.2$ dex. Secondly, and far more important, $\xi$ is explicitly dependent on evolutionary synthesis models, whose predictions for the ionizing fluxes from old stellar populations are still plagued by systematic uncertainties (Section \ref{sec:SSPs}, see also 
\citetalias{Stasinska.etal2008}).  Hence, while it is clear that RGs should populate the bottom of the $\xi$-scale, theory alone does not allow us to establish a firm upper limit below which a galaxy can confidently be considered as retired.  A last caveat is that the evaluation of $\xi$ requires a stellar population analysis, which is not always available or possible.

Fortunately, there is a simple empirical alternative to circumvent these difficulties.

As seen in Fig.\ \ref{fig:xi_X_WHa}a, and in agreement with previous studies (e.g, \citealt{Baumford.etal2008}), the distribution of $W_{\Ha}$ is strongly bimodal, with a low peak centred at $W_{\Ha} = 1$ \AA, an upper one at 16 \AA, and an intermediate minimum in the neighbourhood of 3 \AA.  This happens both for the sample as a whole and for the sub-set of AGN-like galaxies, shown by the filled histogram.  This bimodality strongly suggests that different mechanisms are at work: Photoionization by HOLMES on the low $W_{\Ha}$ side and by massive stars and/or AGN for large $W_{\Ha}$. One can therefore use the observed $W_{\Ha}$ distribution to set an empirical bound for RGs.  Of course, the same can be done for the $\xi$ distribution, but it is clearly advantageous to draw this boundary on the basis of an universally available and model-independent quantity: $W_{\Ha}$.

We conducted various experiments to define an optimal boundary to separate the two peaks, all giving results between 3 and 4 \AA. We settled for $W_{\Ha} = 3$ \AA.  We thus propose to the following practical definition: {\em RGs are ELGs with $W_{\Ha} < 3$ \AA.}

For $W_{\Ha} < 3$ \AA, the central black hole may well be active (i.e., accreting), but its ionizing photon output is comparable to or weaker than that produced by HOLMES.  For $W_{\Ha}^{\rm old}$ values between 1 and 2 \AA, which are compatible with current models (bottom panels in Fig.\ \ref{fig:SSPs}), one finds that at our proposed $W_{\Ha} = 3$ \AA\ borderline the AGN contributes between $\sim 1/3$ and 2/3 of the ionizing power.  Naturally, this fraction decreases even more as $W_{\Ha}$ decreases, to the point that AGN contribution is negligible for the bulk of the low $ W_{\Ha}$ population. It is therefore not correct to use the emission lines of $W_{\Ha} < 3$ \AA\ systems to infer AGN properties.

\section{A comprehensive classification of galaxies}
\label{sec:Revised_EWHan2}

\begin{figure}
\includegraphics[bb= 50 160 580 690,width=0.5\textwidth]{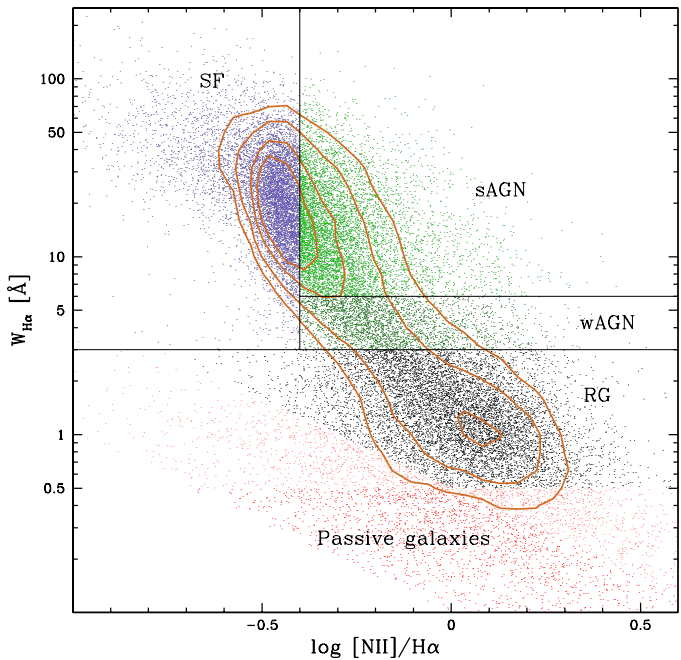}
\caption{The WHAN diagram with the revised categories: SF, sAGN, wAGN,
  RGs and PGs. Points with one of \nii or \Ha weaker than 0.5 \AA\ are
  plotted in gray, except for PGs, which are plotted in red.}
\label{fig:Revised_EWHan2}
\end{figure}

We are now able to separate fake AGN (= RGs) from true AGN.  In the LINER zone of Fig. 1, true AGN are defined by 3 \AA $ < W_{\Ha} < 6$ \AA.  To avoid confusion, we shall call these sources ``weak AGN'' (wAGN). In the interest of a consistent notation, we shall also rename Seyferts as ``strong AGN'' (sAGN). Finally, we remove from the SF category those galaxies consistent with a RG classification, that is to say those having $W_{\Ha} < 3$ \AA.

Our final classification scheme is shown in Fig.\ \ref{fig:Revised_EWHan2}. Sample V sources split into percentage fractions of 22, 21, 8, 31 and 18 SF, sAGN, wAGN, RG and PG, respectively. RGs therefore exist in large numbers in the SDSS, in agreement with the basic prediction that such systems are bound to exits as a mere consequence of stellar evolution.  AGN, on the other hand, are far less common than one would infer associating all LINERs to non-stellar activity.

This revised classification scheme has three main virtues: 

\begin{enumerate}

\item It identifies the three main ionizing agents in galaxies: young stars, AGN and HOLMES.

\item It is based on the cheapest and therefore most inclusive diagnostic diagram, the only one able to handle the huge population of weak line galaxies \citepalias{CidFernandes.etal2010}. In fact, it even allows us to visualize a region occupied by PGs, something which is not possible in conventional intensity ratio diagnostic diagrams (of course, galaxies with no emission lines at all do not appear in this diagram).

\item The SF/AGN and wAGN/sAGN frontiers are based on optimal transpositions of widely used demarcation lines defined on the basis of traditional diagnostic diagrams, such that the WHAN diagram preserves traditional classification schemes where they work well, while at the same time fixing their inability to deal with weak line galaxies and to distinguish true from fake AGN.

\end{enumerate}

Before exploring the realm of galaxies under the perspective of this new classification scheme, a few clarifications are in order.

  First, notice that our SF/AGN border line corresponds to the transposition of the \citetalias{Stasinska.etal2006} BPT-based criterion to isolate systems powered  exclusively  by young stars from those where a harder ionizing source is required. As such, it would have been better to call it a ``pure-SF'' demarcation line, instead of a ``SF/AGN'' divisory line. This is all the more true now that we realize that the AGN-like population of ELGs in the BPT is full of retired galaxies whose black holes are not active enough to affect their emission lines.

  Secondly, notice that the $W_{\Ha} < 3$ \AA\ criterion slightly modifies our idealized definition of RGs as ``systems whose old stellar populations suffice to explain the observed emission line properties'' (Section \ref{sec:XI_model}) to one where HOLMES make at least a substantial contribution to the ionizing field. Such a fuzzy frontier is unavoidable. Our criterion is designed to avoid RGs being misclassified as AGN.  On the other hand, some galaxies hosting an AGN,  particularly those containing a weakly active nucleus, may drift to the $W_{\Ha} < 3$ \AA\ zone if observed within an aperture larger than the SDSS $3^{\prime\prime}$  spectroscopic fiber.  

 Numerical experiments show that this is indeed likely to happen. Picking the \Ha luminosity of a randomly chosen AGN and adding it to the spectrum of a random {\em bona fide} RG with $W_{\Ha}$ within $1 \pm 0.5$ \AA, we find that about 1/3 of the resulting simulated spectra have $W_{\Ha} < 3$ \AA, most of which in the 2--3 \AA\ range. We thus expect that a fraction of the systems here classified as RGs turn out to harbour weak AGN when looked up more closely. At first sight, this seems to be a drawback of using $W_{\Ha}$ for spectral classification, but it is not. After all, if an AGN is present and yet $W_{\Ha} < 3$ \AA, then it is still correct to infer that old stars make a substantial contribution to the observed line emission.

Finally, we note that the division of AGN into strong and weak categories on the basis of $W_{\Ha}$ is also to some extent aperture dependent, since sAGN may be diluted to wAGN as more stellar continuum is sampled.  This, however, is a minor issue, since both classes are clearly AGN-dominated systems, so there is no risk of misidentifying the dominant ionizing agent.

\section{The star formation histories within our classification scheme}
\label{sec:SFHs}

A popular approach to digest the flood of information from mega-surveys like the SDSS is to compare the properties of galaxies in different classes (SF vs.\ AGN, Seyfert vs.\ LINER, early vs.\ late type, field vs.\ cluster, etc.), in search of clues which may help understanding why galaxies are the way they are (see \citealt{Blanton.Moustakas2009} for examples of studies following this general strategy).  Taxonomy, of course, plays a central role in such comparative studies.  In this section we present yet another comparative study, but this time following the new taxonomical framework proposed in the previous section and summarized in Fig.\ \ref{fig:Revised_EWHan2}.  The main novelty in this scheme is the introduction of the RG category, whose members were previously misclassified as AGN.  A major goal of this section is to place RGs in the context of the galaxy population in general, which was never done before.  On the one hand, we want to compare RGs with other ELGs, especially wAGN, with which RGs were so far confused. It is also interesting to compare wAGN to sAGN, since this division was merely introduced for the sake of ``backwards compatibility'' with the historical Seyfert/LINER sub-division of AGNs.  On the other hand, it is equally important to compare RGs and PGs.  According to the RG concept, galaxies without emission lines should {\em not exist}, since HOLMES provide a minimum floor ionizing field capable of powering emission lines. Seen from this perspective, the existence of PGs raises questions like: Are they truly line-less systems or is it just an issue of detectability?  Perhaps PGs lack the warm/cold ISM to reprocess ionizing photons?  

The {\sc starlight}-SDSS database offers plenty of material for this comparative study. In what follows we use a suite of physical properties as well as the detailed star formation histories (SFHs) to address the issues raised above. Only the volume limited sample is used from here on.

\begin{figure}
\includegraphics[bb= 50 155 580 690,width=0.5\textwidth]{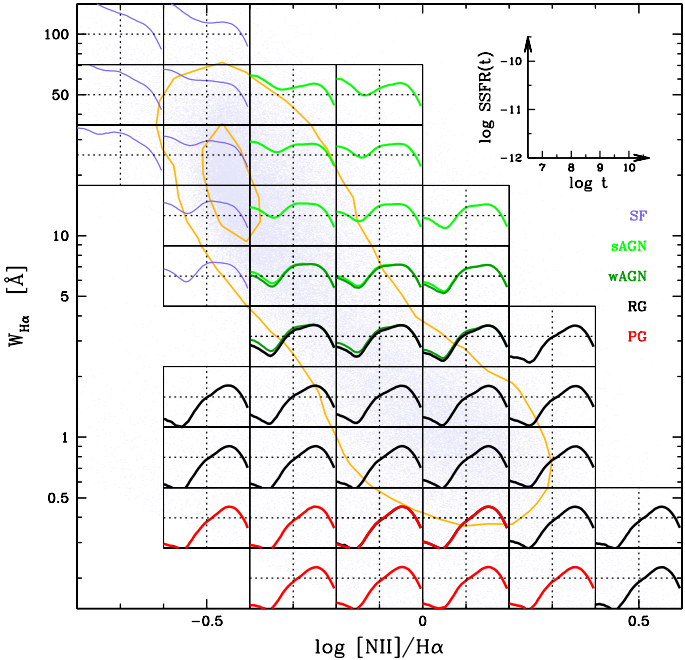}
\caption{Star formation histories as a function of spectral class and
  the location in the WHAN diagram. Each box shows the median
  SSFR(t) for galaxies of different types, coded by different
  colours. The scale of these plots is shown in the inset. Curves are
  only drawn for boxes containing over 200 galaxies.  Only galaxies in
  the volume limited sample are used in this plot.  Contours
  correspond to 0.1 and 0.6 of the peak density.}
\label{fig:SFHs_on_EWhan2}
\end{figure}

An instructive way to start this comparative study is to investigate how SFHs vary as a function of the SF/sAGN/wAGN/RG/PG spectral types defined in Fig.\ \ref{fig:Revised_EWHan2}.  This is done in Fig.\ \ref{fig:SFHs_on_EWhan2} where we chopped the WHAN diagram in $0.2 \times 0.3$ dex wide bins in \nii/\Ha and $W_{\Ha}$. In each box we plot the time-dependent median of the specific star formation rate (SSFR) for galaxies in the box, colour coding the curves according to the Fig.\ \ref{fig:Revised_EWHan2} emission line taxonomy.  The SSFR(t) functions were smoothed exactly as in \citet{Asari.etal2007}, which should be consulted for details on how this and other representations of {\sc starlight}-based SFHs are derived.

Because this is our classification diagram, most boxes contain only one curve, with exceptions at locations straddling the nominal frontier between RGs and wAGN (at $W_{\Ha} = 3$ \AA) and wAGN/sAGN ($W_{\Ha} = 6$ \AA). Although there are strong systematic changes in SFHs across this plane, in these border-line cases one sees little difference between the SFHs of two adjacent classes.

Several trends are visible in Fig.\ \ref{fig:SFHs_on_EWhan2}.  For instance, focusing on the column centred at $\log \nii/\Ha = -0.1$, one sees that both PGs and RGs have SFHs concentrated in the largest ages, while, moving upwards the retrieved SFHs reveal increasing levels of recent star-formation. Due to the large apertures covered by the SDSS fibers, most of the \Ha emission in these AGN probably comes from star-formation rather than the AGN \citep[e.g.][]{Kauffmann.etal2003c}, so the increase in recent SSFR with $W_{\Ha}$ is expected. However, the effect of the AGN is present, as can be inferred by the decreasing levels of recent SSFR towards the right for fixed $W_{\Ha}$ (best visible in the rows centred at $W_{\Ha} \sim 10$ \AA).

\begin{figure}
\includegraphics[bb= 50 155 580 690,width=0.5\textwidth]{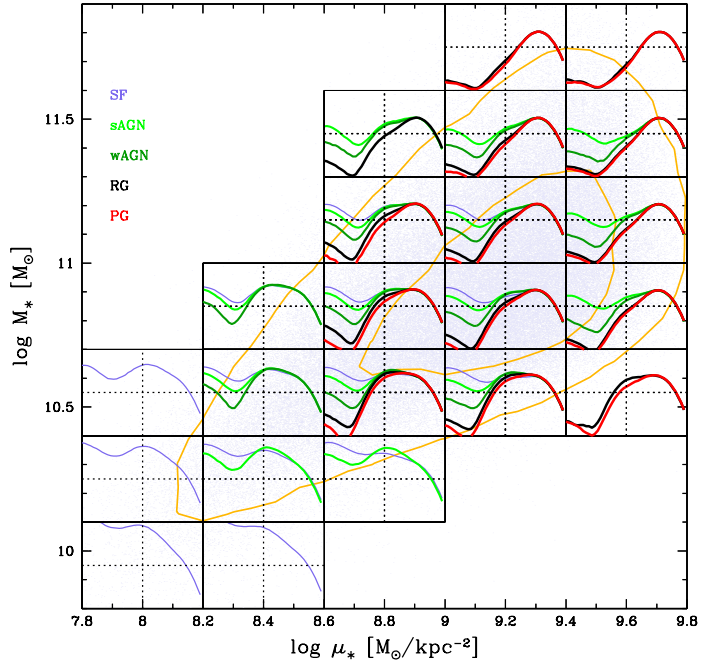}
\caption{As Fig.\ \ref{fig:SFHs_on_EWhan2}, but in the plane defined
  by the mean stellar surface density and stellar mass. $\mu_\star$ is
  computed dividing half the stellar mass by the radius which contains
  half the z-band light.  }
\label{fig:SFHs_on_M_x_SMD}
\end{figure}

Before we proceed, a parenthesis must be opened to clarify the meaning of the small upturn in SSFR(t) at $t \la 10^7$ yr for RGs and PGs.  This feature is almost certainly an artifact of incomplete modelling of the horizontal branch phase \citep{Koleva.etal2008, Ocvirk2010, CidFernandes.GonzalezDelgado2010}.  This effect was quantified by \citet{Ocvirk2010}, who estimates that these ``fake young bursts'' account for up to 12\% of the recovered optical light fractions in spectral synthesis of genuinely old stellar populations. In units of the SSFR, this corresponds to a few times $10^{-12}\,$yr$^{-1}$, which is precisely the level at which upturns at small ages are seen in Fig.\ \ref{fig:SFHs_on_EWhan2}. We further find a clear trend for this artefact to be related to the stellar metallicity, with the low $t$ hump increasing for decreasing $\overline{Z_\star}$, in agreement with the fact that bluer horizontal branches occur for lower $Z_\star$ (\citealt{Harris1996}; see also Fig.\ \ref{fig:SFHs_on_M_x_SMD} for indirect evidence of this trend). Finally, if the upturn were really due to young stars, $W_{\Ha}$ should be much larger than observed. Including these fake young bursts in the predictions for the ionizing flux leads to (in the median) expected $W_{\Ha}$ values over 3 times as large as observed for RGs. The overprediction for PGs is even worse, with a median expected $W_{\Ha}$ of 3.2 \AA, while their definition requires $W_{\Ha} < 0.5$ \AA.

Credible levels of recent star-formation only appear for boxes centred at $W_{\Ha} \ge 3$ \AA\ in Fig.\ \ref{fig:SFHs_on_EWhan2}, coinciding with the RG/wAGN transition. This suggests that {\em all} AGN are associated to at least some level of ongoing star-formation. While this is well known and documented for sAGN (the so called ``starburst-AGN connection'') the situation is much less clear for wAGN (e.g. \citealt{GonzalezDelgado.etal2004} and references therein). It is difficult to compare this finding with those in previous studies because of varying definitions of wAGN, but we can confidently say that the signatures of star formation in wAGN identified in this work would be wiped out without removing the RGs from the AGN category.

We now investigate how SFHs vary in terms of properties not explicitly involved in the definition of our spectral classes.  Fig.\ \ref{fig:SFHs_on_M_x_SMD} shows SFHs in a stellar mass ($M_\star$) versus mass density ($\mu_\star$) diagram.  This allows us to compare the SFHs of galaxies which are similar in terms of global properties, but differ in terms of emission line properties, since this time each box can contain galaxies of different classes. Again, several trends are evident to the eye. One of them is the signature of downsizing, here manifested by the decrease in recent SSFR levels as one moves up the mass scale, seen, for instance, in the curves for SF galaxies in the 1$^{\rm st}$ two columns of boxes.  A similar behaviour is also seen for RGs and PGs\footnote{Recall that the ``mini bursts'' at $t \la 10^7 $ yr are not real, but artifacts of spectral fits with an incomplete base \citep{Ocvirk2010}. The fact that these fake bursts correlate with galaxy mass, as seen in Fig.\ \ref{fig:SFHs_on_M_x_SMD}, is likely an indirect consequence of the connection between horizontal branch morphology and $Z_\star$ coupled to the $M_\star$--$Z_\star$ relation.}, but shifted in time to $t \sim 10^9$ yr.
 
For this study, the more relevant trend in Fig.\ \ref{fig:SFHs_on_M_x_SMD} is that in every box the lines corresponding to our different classes appear in the same order, with recent star formation decreasing from SF galaxies to sAGN, wAGN, RGs and PGs. These last two, in fact, have median SFHs which are very similar and sometimes indistinguishable. Both types retired from forming stars over $10^8$ yr ago.  The plot also shows that the recent SSFR of wAGN never reaches levels as low as those of RGs and PGs. Despite their generally low levels of star formation compared to sAGN of same $M_\star$ and $\mu_\star$, wAGN are not retired, reinforcing the idea that star formation and AGN activity are connected even in the weakest cases, at least in a statistical sense.

\begin{figure}
\includegraphics[bb= 50 155 580 690,width=0.5\textwidth]{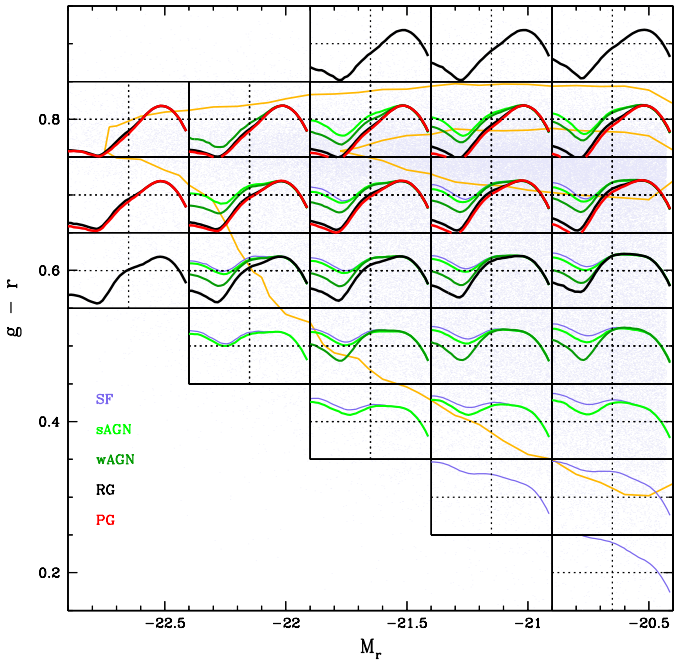}
\caption{As Fig.\ \ref{fig:SFHs_on_EWhan2}, but for the observational color magnitude diagram.}
\label{fig:SFHs_on_2in1}
\end{figure}

Fig.\ \ref{fig:SFHs_on_2in1} shows another example. The plot shows the observational color magnitude diagram, with the familiar red sequence, blue cloud and green valley pattern. This diagram is meant to provide an empirical representation of the age-mass relation.  As expected, systems in the blue cloud are more actively forming stars nowadays than in their past average, while red sequence galaxies (with the exception of AGN) tend to be retired.  Again, the SF to PG sequence in levels of recent SSFR and the similarity between RGs and PGs seen in Figs.\ \ref{fig:SFHs_on_EWhan2} and \ref{fig:SFHs_on_M_x_SMD} is present in every box.

In summary, the extended emission line classification scheme proposed here bears an excellent correspondence with the SFHs of galaxies.  RGs and PGs have very similar SFHs for similar global properties.  wAGN are more similar to sAGN in terms of SFHs than to RGs or PGs.  Traditional classification schemes which ignore RGs, would, on the contrary, find wAGN to be more like PGs.

\section{Distributions of global galaxy properties}
\label{sec: distributions}

\begin{figure*}
\includegraphics[bb= 50 175 580 620,width=\textwidth]{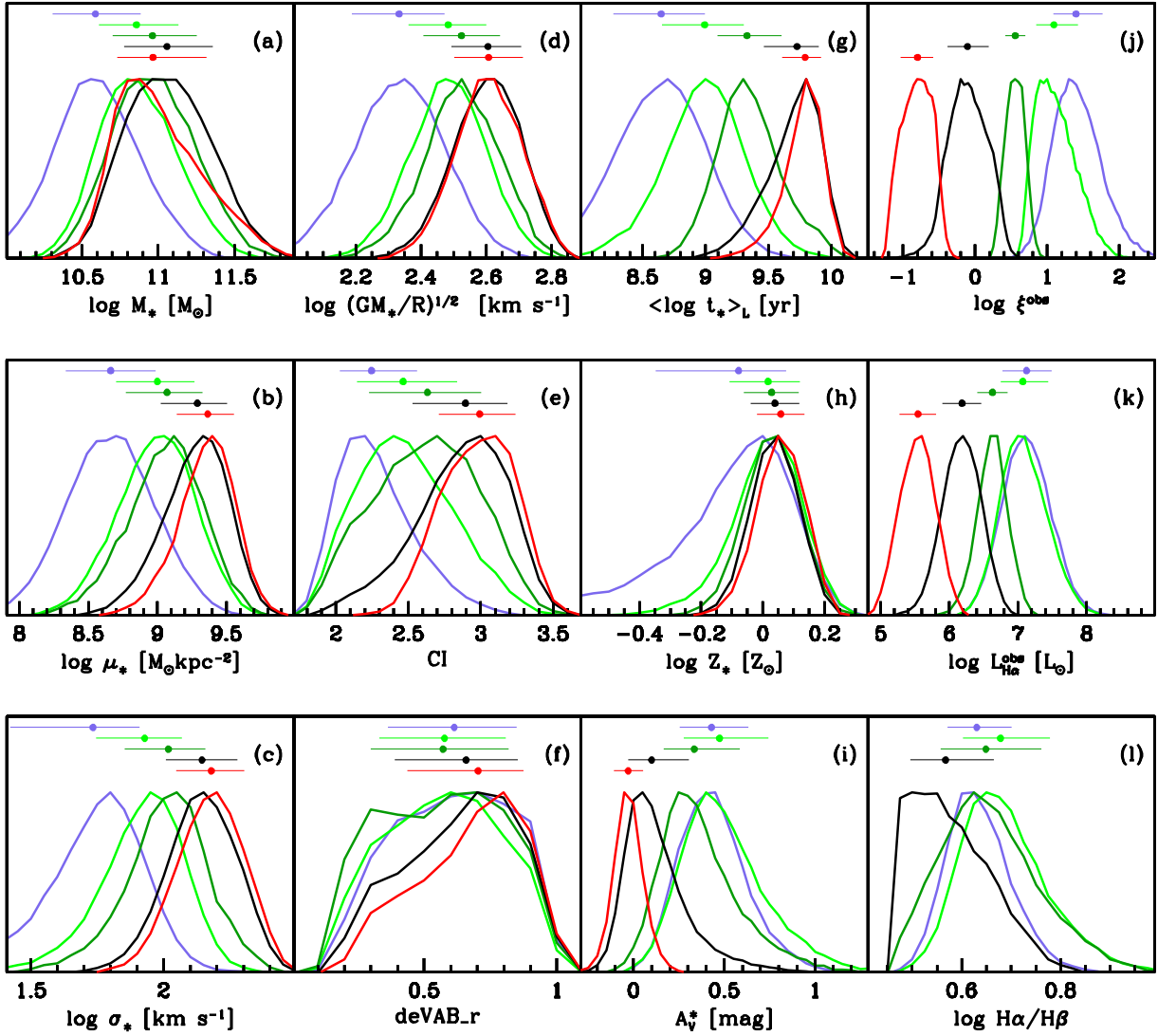}
\caption{Normalized histograms of properties for different galaxy
  classes. Colours code the SF, sAGN, wAGN, RG and PG classes as in
  Fig.\ \ref{fig:Revised_EWHan2}.  Horizontal bars trace the 16 to 84
  percentile ranges, with the median marked by a circle, with the
  vertical shifts ordered as SF (top), sAGN, wAGN, RG and PG
  (bottom). All galaxies from our volume limited sample are included,
  except for panel l, for which a reliable \Ha/\Hb was required, hence
  excluding all PGs and most RGs. }
\label{fig:HistsProps}
\end{figure*}

Comparing the distributions of global properties of different galaxy classes can provide a further test of the pertinence of our classification scheme.  This is done in Fig.\ \ref{fig:HistsProps}, which shows normalized histograms of PG, RG, wAGN, sAGN and SF galaxies in terms of a series of properties related to their structure, stellar populations, gas and dust content.

A striking feature of these histograms is that for most properties PG and RG have similar distributions, reinforcing our interpretation that PG and RG are essentially the same thing, or at least that these classes overlap heavily. On the other hand, the distributions for wAGN are more similar to the ones of sAGN, suggesting that these galaxies belong to the same family.  This is a salutary a posteriori validation of our splitting of LINERs into two very different classes: RGs and wAGN.

Except for SF galaxies, all other galaxy types have masses within roughly $\pm 0.5$ dex of $10^{11}$ M$_\odot$ (Fig.\ \ref{fig:HistsProps}a). This is mostly due to the luminosity cut for the V sample, which imposes a mass completeness limit of $\sim 10^{10.8} M_{\odot}$. In the median, $M_\star^{\rm sAGN} < M_\star^{\rm wAGN} \sim M_\star^{\rm PG} < M_\star^{\rm RG}$.  In terms of surface densities and velocity dispersions (panels b and c), PGs and RGs are statistically similar, occupying the high end of the distributions. Despite the substantial overlap, it is clear that wAGN and sAGN tend to have smaller $\sigma_\star$ and $\mu_\star$ than RGs and PGs.  The same is true for the escape velocity, computed from $(GM_\star/R)^{1/2}$ with $R$ given by the half-light z-band radius (panel d).  PGs and RGs also differ from AGN in terms of structural properties like the concentration index (e) and the SDSS deVAB\_r index (f). Inspection of SDSS images corroborates the similarities between PGs and RGs, most of which are ellipticals, although some of the RGs turn out to be spirals (as can be deduced by the tail towards low CI in Fig.\ \ref{fig:HistsProps}e), including an apparently higher than normal rate of edge-on systems.  An investigation of this issue is beyond the scope of this paper, but we note that this ``contaminant'' population tends to have $W_{\Ha}$ close to the RG/wAGN frontier, and thus may be related to the unavoidable fuzziness of classification schemes.

In terms of luminosity-weighted mean stellar ages (Fig.\ \ref{fig:HistsProps}g), sAGN are younger than wAGN, which in turn are younger than RGs and PGs.  From Fig.\ \ref{fig:SFHs_on_EWhan2} we infer that the tail of RGs towards smaller mean ages is due to systems at the RG/wAGN border.  The mean stellar metallicities (panel h) are similar for all types, with a tail of sAGN reaching lower values.

The distributions of stellar extinctions (panel i) show a distinct PG $<$ RG $<$ wAGN $<$ sAGN pattern. To double check this result, panel l shows the distribution of \Ha/\Hb, but only for sources where this ratio is reliable, a criterion which excludes PGs and most RGs.  Despite its incompleteness, Fig.\ \ref{fig:HistsProps}l confirms that wAGN have higher extinction than RGs and lower than sAGN. The class with the lowest extinction is the SF class, in spite of the intense star formation which is usually associated with high extinction. The underlying reason for a smaller extinction in SF than the rest of galaxies lies in their lower masses \citep[see ][]{Stasinska.etal2004, Garn.Best2010}. Regarding the PG $\rightarrow$ RG $\rightarrow$ wAGN $\rightarrow$ sAGN extinction sequence, its interpretation is not straightforward, since it involves the dust mass and distribution, as well as its composition (dust grains recently produced by the progenitors of HOLMES and expected to be found in RGs do not have the same extinction properties as grains processed in the interstellar medium and associated with regions of star formation).

Fig.\ \ref{fig:HistsProps}j shows the distribution of $\xi$ uncorrected for dust. The PG $\rightarrow$ RG $\rightarrow$ wAGN $\rightarrow$ sAGN $\rightarrow$ SF sequence reflects the central role of $W_{\Ha}$ in our classification scheme. This sequence is mirrored in the distribution of observed \Ha\ luminosities (Fig.\ \ref{fig:HistsProps}k). PGs included in these panels have, by definition, very weak \Ha (median $W_{\Ha} = 0.23$ \AA, and median $SN_{\Ha} = 2.4$ counting only those with an $\Ha$ flux in our database). Still, their locations in these diagrams are consistent with them not being completely line-less.

\section{Discussion and summary}
\label{sec:summary}

The goal of this paper was to provide a comprehensive classification of galaxies according to their emission line properties which would be able to cope with the large population of weak line galaxies that do not appear in traditional diagrams (e.g., BPT) because they lack some of the diagnostic lines. This is possible using the WHAN diagram ($W_{\Ha}$ vs \nii/\Ha) that we proposed in earlier work \citepalias{CidFernandes.etal2010}. An additional problem with the BPT diagram and other similar line-ratio diagrams is that, as shown by \citet{Stasinska.etal2008}, the LINER region contains a mix of two completely different families of galaxies: galaxies that contain a weak AGN and ``retired galaxies'', i.e. galaxies that have stopped forming stars and whose emission lines are powered by their hot evolved low-mass stars (HOLMES).  A truly comprehensive classification scheme must account for this phenomenon, and the WHAN diagram allows so.

To distinguish true AGN from fake ones (i.e., the RGs), we have shown that a useful criterion is the value of $\xi$, which measures the intrinsic extinction-corrected \Ha\ luminosity in units of the \Ha\ luminosity expected from stellar populations older than $10^{8}$\,yr uncovered by our stellar population analysis software {\sc starlight}.  In principle, $\xi$ should be equal to one if the galaxy absorbs all the ionizing photons provided by its HOLMES, and $> 1$ if any other source of ionization is present.  We have shown that this index is bimodally distributed, with a $\xi \gg 1$ population composed by galaxies undergoing vigorous star-formation and/or nuclear activity, and a population of galaxies centred right at the predicted value for RGs. Computing $\xi$ for each galaxy, however, is not a trivial task: it requires a stellar population analysis able to identify the populations of HOLMES and a recipe to obtain the ionizing radiation field from these HOLMES. The latter strongly depends on how the (yet quite uncertain) evolutionary tracks for post-asymptotic giant branch stars are incorporated into the stellar population codes as well as on other issues such as the IMF and the metallicities of the stellar populations.

It is therefore preferable to base the distinction between true and fake AGN on the \Ha\ equivalent width, an easy to obtain observational parameter which provides an excellent proxy for $\xi$. In view of the strong dichotomy in the observed $W_{\Ha}$ distribution for galaxies in the right wing of the BPT diagram (i.e., the wing previously supposed to contain only AGN), we propose a practical separation between weak AGN and RGs based on a class frontier at $W_{\Ha}=3$ \AA.  This physically inspired but data-guided separation corresponds to the minimum between the two modes in the $W_{\Ha}$ distribution.

We thus identify 5 classes of galaxies in the WHAN diagram: 

\begin{itemize}
  \item SF: Pure star forming galaxies: log \nii/\Ha\ $< -0.4$ and  $W_{\Ha} > 3$ \AA
  \item sAGN: Strong AGN (i.e., Seyferts): log \nii/\Ha\ $> -0.4$ and  $W_{\Ha} > 6$ \AA
  \item wAGN: Weak AGN: log \nii/\Ha\ $> -0.4$ and  $3  <W_{\Ha} < 6$ \AA
  \item RG: Retired galaxies (i.e., fake AGN): $W_{\Ha} < 3$ \AA
  \item PG: Passive galaxies (actually, line-less galaxies): $W_{\Ha}$ and $W_{\nii} < 0.5$ \AA
\end{itemize}

Note that the border between RGs and PGs is somewhat arbitrary, but this is not really a concern.

We have then analysed the star formation histories of the different classes of galaxies, as well as the distribution of a dozen of physical or observational properties, such as the galaxy stellar mass $M_\star$, surface densities $\mu_\star$, velocity dispersions $\sigma_\star$, and stellar extinctions $A_V^\star$ obtained through {\sc starlight}.  All these comparisons corroborate our proposed differentiation between RGs and wAGN in the LINER-like family.  We also find that wAGN have properties overlapping with those of sAGN, while RGs line up with PGs.  In other words, wAGN are the low activity end of the AGN family, while RGs are passive galaxies with emission lines.  As a matter of fact, both RGs and PGs are ``retired'' or ``passive'' galaxies, the real spectroscopic difference between them being the presence or absence of emission lines. It is only the inherited custom which imposed the adopted nomenclature.

Now that these emission line classes have been established, further work could be to understand what is at the root of the strength of the AGN phenomenon  or why some passive galaxies present emission lines and others do not.


\section*{ACKNOWLEDGEMENTS}

The {\sc starlight} project is supported by the Brazilian agencies CNPq, CAPES, by the France-Brazil CAPES-COFECUB program, and by Observatoire de Paris.

Funding for the SDSS and SDSS-II has been provided by the Alfred P. Sloan Foundation, the Participating Institutions, the National Science Foundation, the U.S. Department of Energy, the National Aeronautics and Space Administration, the Japanese Monbukagakusho, the Max Planck Society, and the Higher Education Funding Council for England. The SDSS Web Site is http://www.sdss.org/.

The SDSS is managed by the Astrophysical Research Consortium for the Participating Institutions. The Participating Institutions are the American Museum of Natural History, Astrophysical Institute Potsdam, University of Basel, University of Cambridge, Case Western Reserve University, University of Chicago, Drexel University, Fermilab, the Institute for Advanced Study, the Japan Participation Group, Johns Hopkins University, the Joint Institute for Nuclear Astrophysics, the Kavli Institute for Particle Astrophysics and Cosmology, the Korean Scientist Group, the Chinese Academy of Sciences (LAMOST), Los Alamos National Laboratory, the Max-Planck-Institute for Astronomy (MPIA), the Max-Planck-Institute for Astrophysics (MPA), New Mexico State University, Ohio State University, University of Pittsburgh, University of Portsmouth, Princeton University, the United States Naval Observatory, and the University of Washington.


\end{document}